\documentclass[twoside]{article}
\usepackage{graphicx}
\usepackage{fancyhdr}
\usepackage{multicol}
\usepackage{styl-ap}

\begin{document}
\title{Mass Accretion Processes in Young Stellar Objects: Role of Intense Flaring Activity}
\author{S. Orlando\work{1}, F. Reale\work{2,1}, G. Peres\work{2,1}, A. Mignone\work{3}}
\workplace{INAF - Osservatorio Astronomico di Palermo, Piazza del Parlamento 1, 90134, Palermo, Italy
\next
Dip. di Fisica e Chimica, Universit\`a degli Studi di Palermo, Piazza del Parlamento, 1, 90134, Palermo, Italy
\next
Dip. di Fisica Generale, Universit\`a degli Studi di Torino, via Pietro Giuria 1, 10125, Torino, Italy}
\mainauthor{orlando@astropa.inaf.it}
\maketitle

\begin{abstract}%
According to the magnetospheric accretion scenario, young low-mass
stars are surrounded by circumstellar disks which they interact
with through accretion of mass. The accretion builds up the star
to its final mass and is also believed to power the mass outflows,
which may in turn have a significant role in removing the excess
angular momentum from the star-disk system. Although the process
of mass accretion is a critical aspect of star formation, some of
its mechanisms are still to be fully understood. On the other hand,
strong flaring activity is a common feature of young stellar objects
(YSOs). In the Sun, such events give rise to perturbations of the
interplanetary medium.  Similar but more energetic phenomena occur
in YSOs and may influence the circumstellar environment. In fact, a
recent study has shown that an intense flaring activity close to
the disk may strongly perturb the stability of circumstellar disks,
thus inducing mass accretion episodes (Orlando et al. 2011).
Here we review the main results obtained in the field and the future
perspectives.
\end{abstract}


\begin{multicols}{2}

\section{Introduction}

Observations in the X-ray band reveal that low-mass pre-main-sequence
stars are strong sources with X-ray luminosities $3-4$ orders of magnitude
greater than that of the present-day Sun. The source of this high-energy
radiation is plasma with temperatures of $1-100$ MK in the stellar outer
atmospheres (coronae), heated by magnetic activity analogous to the
solar one but higher by factors up to $10^6$. Such a magnetic activity
manifests through very different phenomena that may occur in several
places of the stellar atmosphere and circumstellar environment. The young
star interacts with its disk in a complex fashion, with accretion and
ejection of collimated outflows. Strong magnetic fields are believed
to connect the star with a Keplerian circumstellar disk, funneling
accretion onto limited portions of the stellar surface (e.g. Hartmann
1998) where shocks are produced by the impact of the accretion streams
(e.g. Orlando et al. 2010).

X-ray flares are violent manifestations of the stellar magnetic
activity and are triggered by an impulsive energy input from coronal
magnetic fields. X-ray observations in the last decades have shown
that flares in young stellar objects (YSOs) have amplitudes much
larger than solar analogues and occur much more frequently. Examples
of these flares are those collected by the Chandra satellite in the
Orion star-formation region (COUP enterprise; Favata et al. 2005).
In the Sun, such energetic events are often associated to coronal
mass ejections and give rise to perturbation effects of the
interplanetary medium, broadly known as space weather effects.
Similar phenomena are expected to occur in young stars, and may
affect the circumstellar environment. For instance, Favata et al.
(2005) analyzed the most energetic flares observed by COUP and found
that these flares might be hosted in magnetic loops extending several
stellar radii, much larger than ever observed in older stars. Since
the central star is surrounded by a circumstellar disk accreting
material onto the star, it is natural to ask whether strong flaring
activity involves the disk and even perturbs its stability, possibly
affecting the mass accretion to the star.

At the present time, in fact, it is unclear where these flares
occur. The differential rotation of the disk together with the
interaction of the disk with the magnetosphere may cause magnetic
reconnection close to the disk's surface, triggering large-scale
flares there. In this case, the flares may perturb the stability
of the circumstellar disk causing, in particular, a strong local
overpressure. The pressure gradient force might be able to push
disk's matter out of the equatorial plane into funnel streams, thus
providing a mechanism to drive mass accretion that differs from
that, commonly invoked in the literature, based on the disk viscosity
(Romanova et al.  2002). Bright flares close to circumstellar disks
may therefore have important implications for a number of issues
such as the transfer of angular momentum and mass between the star
and the disk, the powering of outflows, and the ionization of
circumstellar disks.

Recently, we have investigated the idea that an intense flare close
to an accretion disk may perturb the stability of the disk and
trigger mass accretion onto the star (Orlando et al. 2011).  In
this paper we review our findings and present preliminary results
of a study investigating the effects of a storm of small-to-medium
flares on the stability of accretion disks (Orlando et al. 2014).
In Sect.  \ref{sec2} we describe the MHD model; in Sect. \ref{sec3}
we present the results; in Sect. \ref{sec4} we draw our conclusions.

\section{The MHD Model}
\label{sec2}

The model describes a classical T Tauri star (CTTS) of mass $M_{*} =
0.8 M_{\odot}$ and radius $R_* = 2 R_{\odot}$ located at the origin of a
3D spherical coordinate system (see Orlando et al. 2011 for a detailed
description). We adopted the initial conditions introduced by Romanova
et al. (2002). In particular, we assumed the rotation period of the
star to be 9.2 days as representative of CTTSs. The initial unperturbed
stellar atmosphere is approximately in equilibrium and consists of three
components: the stellar magnetosphere, the extended stellar corona,
and the Keplerian disk.  Initially, the magnetosphere is assumed to be
force-free, with dipole topology and magnetic moment aligned with the
rotation axis of the star. The magnetic moment is chosen in order to have
a magnetic field strength of the order of 1 kG at the stellar surface.
The isothermal disk is cold, dense and rotates with angular velocity
close to the Keplerian value; its rotation axis (coincident with the
rotation axis of the star) is aligned with the magnetic moment. The disk
is initially truncated by the stellar magnetosphere at the radius $R_d$
where the ram pressure of the disk is equal to the magnetic pressure;
for the adopted parameters, $R_d = 2.86 R_*$ and the co-rotation radius
is located at $R_{co} = 9.2 R_*$. The corona is initially isothermal
with temperature $T = 4$ MK and at low density.

The system is described by the time-dependent MHD equations of mass,
momentum, and energy conservation in a 3D spherical coordinate
system, extended with gravitational force, viscosity of the Keplerian
disk, thermal conduction (including the effects of heat flux
saturation), coronal heating (via a phenomenological term), and
radiative losses from optically thin plasma (see Orlando et al.
2011 for more details). The phenomenological heating is prescribed
as a component, describing the stationary coronal heating, plus a
transient component, triggering the flares. The calculations were
performed using PLUTO (Mignone et al. 2007), a modular, Godunov-type
code for astrophysical plasmas.

\section{Results}
\label{sec3}

As a first application of the model, we have investigated the effects
of a single bright flare on the stability of the disk (Orlando et al.
2011). The initial heat pulse triggering the flare is released at a
distance of $5 R_*$, namely in a region comprised between the truncation
and corotation radii, and it is supposed to be indicative of a likely
area of magnetic reconnection. Heat pulses occurring closer to the
inner disk edge are expected to produce analogous perturbations on the
disk dynamics. We followed the evolution of the star-disk system for
$\approx 2$ days. After the heat pulse has been released, an MHD shock
wave develops above the disk and propagates away from the protostar. The
heat deposition determines a local increase of temperature and pressure.
Disk material is heated and expands with a strong evaporation front
which is channeled along the magnetic field lines towards the central
star. After $\sim 1$ hour a hot loop forms linking the disk with the star
(see upper panel in Fig.~\ref{fig1}). The loop has an effective maximum
temperature of $\approx 100$ MK (at the peak of emission measure) and
a length of $\approx 10^{12}$~cm; these values are in good agreement
with those inferred from the analysis of the brightest flares observed
by COUP (Favata et al. 2005). The overheating of the disk surface
makes a significant amount of material expand and be ejected in the
magnetosphere. A small fraction of this material fills the loop whose
density increases from $10^8$~cm$^{-3}$ to $10^{10}$~cm$^{-3}$.  On the
other hand, most of the expanding disk material is not confined by the
magnetic field and is ejected away from the star, carrying away mass
and angular momentum.

During the evolution of the hot loop, an overpressure wave develops
where the heat pulse has been injected.  This overpressure travels
through the disk and reaches the opposite boundary after $\approx 5$
hours. There the pressure gradient force drives the material out of the
disk and channels it into a funnel flow. Then the gravitational force
accelerates the escaped material toward the central star where the stream
impacts $\approx 25$ hours after the injection of the heat pulse. The
accretion flow persists until the end of the simulation ($t = 48$ h).
The lower panel in Fig.~\ref{fig1} shows a cutaway view of the
star-disk system after the impact of the stream onto the stellar surface.
We found that the dynamics and physical characteristics of the accretion
stream triggered by the flare closely recall those of streams driven by
the accumulation of mass at the disk truncation radius under the effect
of the viscosity and pushed out of the equatorial plane because of the
growing pressure gradient there (e.g. Romanova et al. 2002).

\begin{myfigure}
\centerline{\resizebox{76mm}{!}{\includegraphics{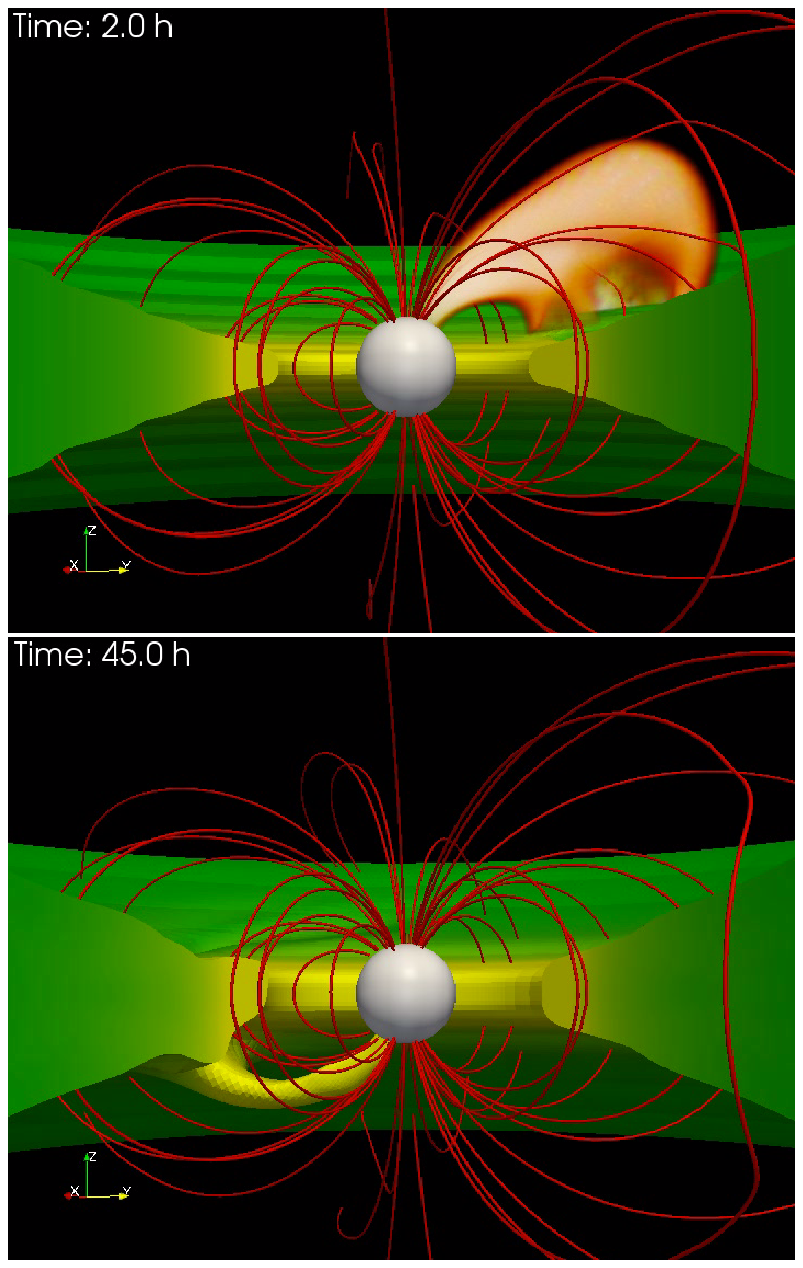}}}
\caption{Effects of a single bright flare on the stability of the
	 circumstellar disk. Cutaway views of the star-disk system
	 showing the mass density of the disk (yellow-green) at
	 $t=2.0$ hours (upper panel) and at $t=45$ hours (lower
	 panel) since the injection of the heat pulse. The upper
	 panel also over-plots the three-dimensional volume rendering
	 of the plasma temperature (MK), showing the flaring loop (in
	 red) linking the inner part of the disk with the star. The
	 lower panel shows the accretion stream triggered by the
	 flare in the side of the disk opposite to the flaring loop.
	 Selected magnetic field lines are overplotted in red.}
\label{fig1}
\end{myfigure}

From the simulation, we derived also the mass accretion rate and
found $\dot{M} > 2.5\times 10^{-10}\,M_{\odot}$~yr$^{-1}$. We
compared this rate with those inferred from optical observations
of CTTSs (Herczeg \& Hillenbrand 2008; Curran et al.  2011) and
found a good agreement.  We concluded that a bright flare as those
frequently detected in YSOs (e.g. COUP observations) can be an
efficient mechanism to trigger accretion onto the protostar itself
with accretion rates on the same order of those commonly measured
in CTTSs.

As a follow-up of the previous study, we explored in more
details the possibility that significant mass accretion in young stars
can be triggered by a storm of small-to-medium flares (as those frequently
observed) occurring on the accretion disk (Orlando et al. 2014).  To this
end, we performed a 3D MHD simulation analogous to that described in
Orlando et al. (2011) but considering a storm of flares distributed
randomly in proximity of the disk surface instead of a single bright
flare.

We found that each simulated flare follows an evolution similar to
that of the single bright flare described in Orlando et al. (2011).
The main difference is that, now, interactions between next flares
may occur. Figure~\ref{fig2} shows cutaway views of the star-disk
system after $\approx 26$ hours. During the system evolution, the
flares build up an extended corona linking the star with the disk.
At the same time, the disk is strongly perturbed by the flares and,
after $\approx 20$ hours, several funnel streams develop, accreting
substantial mass onto the star (see lower panel in Fig.~\ref{fig2}). The
streams persist until the end of the simulation and last for a time longer
than the typical interval between flares.  The simulated mass accretion
rate is $10^{-10} < \dot{M} < 10^{-9} \,M_{\odot}$˜yr$^{-1}$, again in
good agreement with the values inferred from optical observations of CTTSs
(Herczeg \& Hillenbrand 2008; Curran et al. 2011).

\section{Conclusions}
\label{sec4}

We investigated the effects of an intense flaring activity on the
stability of a circumstellar disk surrounding a magnetized CTTS.
To this end, we developed a 3D MHD model including, for the first
time, all key physical processes, most notably the thermal conduction
and the radiative losses from optically thin plasma. As a first
step, we analyzed the perturbation induced by a single bright flare
occurring in proximity of the disk (Orlando et al. 2011). Then, we
investigated the effects of a storm of small-to-medium flares
distributed randomly close to the disk (Orlando et al. 2014). Our
findings lead to the following conclusions: (a) flares occurring close to
the circumstellar disk can trigger substantial and persistent accretion
flows, similar to those caused by the disk viscosity; (b) an intense
flaring activity close to the disk builds up an extended corona linking
the star to the disk.

In the case of a single bright flare occurring close to the disk,
the simulations suggest that mass accretion events associated with
X-ray flares should be observed. However, correlation between UV/optical
accretion tracers and X-ray flux is rarely seen (Stassun et al. 2006). On
the other hand, such a correlation is not prediced by simulations
describing a continuous flaring activity close to the disk (Orlando et
al. 2014). In this case, the streams are triggered by the first flares
at the beginning of the simulation and then are continuously powered by
the following ones. At regime, therefore, no clear correlation between
UV/optical accretion tracers and X-ray flux is foreseen. In the light
of the above findings, we suggest that the flaring activity common to
YSOs may turn out to be important in the exchange of angular momentum
and mass between the circumstellar disk and the central protostar.

\begin{myfigure}
\centerline{\resizebox{76mm}{!}{\includegraphics{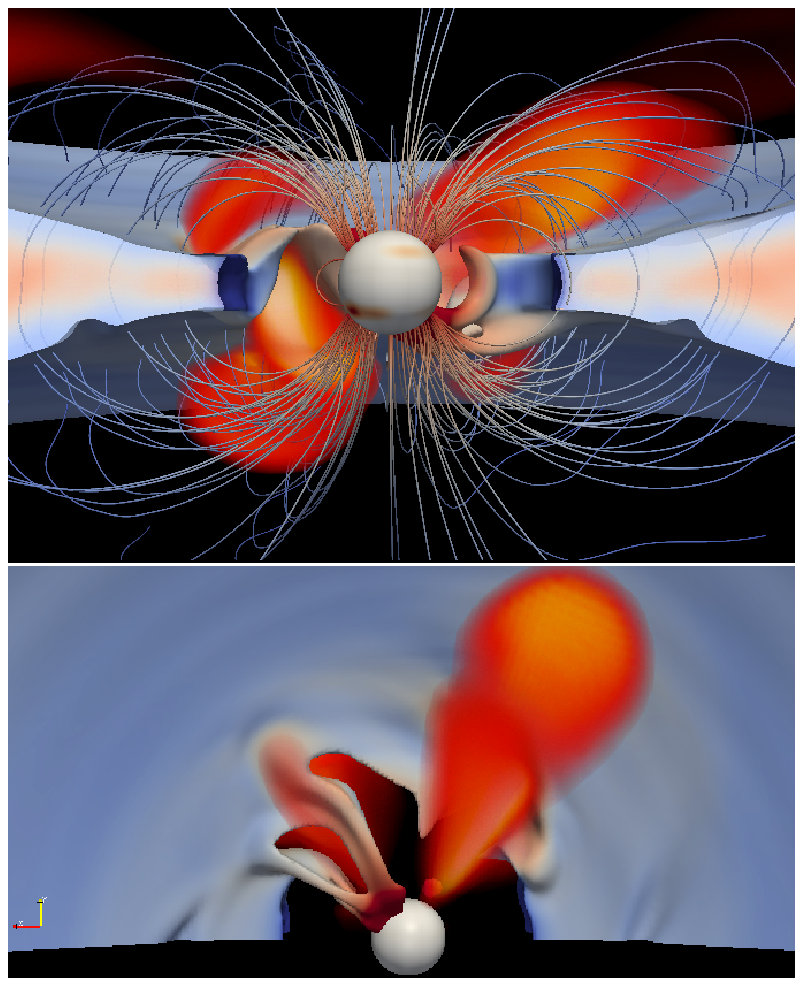}}}
\caption{Effects of a storm of flares on the disk stability. Cutaway
	 views of the star-disk system showing the mass density of
	 the disk (blue) at $t=26$ hours. The panels over-plot the
	 three-dimensional volume rendering of the plasma temperature
	 (in MK), showing the flaring loops (in red) linking the
	 inner part of the disk with the star. The upper panel shows
	 the star-disk system observed edge-on, whereas the lower
	 panel shows the system observed pole-on. Selected magnetic
	 field lines are overplotted in the upper panel.}
\label{fig2}
\end{myfigure}

\thanks
We thank the referee for constructive and helpful criticism.
PLUTO is developed at the Turin Astronomical Observatory in
collaboration with the Department of General Physics of the Turin
University. We acknowledge the CINECA Award N. HP10A4ZCV5,2012 for
the availability of high performance computing resources and support.

\bigskip
\bigskip
\noindent {\bf DISCUSSION}

\bigskip
\noindent {\bf MATTEO GUAINAZZI:} What kind of X-ray variability
does your model predict? is it observed in the COUP sources?

\bigskip
\noindent {\bf SALVATORE ORLANDO:} The X-ray variability depends
on the rate of flares triggered. In the simulation with a single
flare, we observe the fast rise and subsequent slow decay of X-ray
emission characteristic of flares. In the simulation with a storm
of flares, we observe a background emission due to the many small
flares evolving simultaneousluy with superimposed local peaks due
to the brightest flares. The X-ray variability found is consistent
with that of YSOs observed by COUP.

\bigskip
\noindent {\bf DAVID BUCKLEY:} Do you believe that your magnetospheric
disk interaction models could be extrapolated to intermediate polar
class of CVs, where accretion occurs from disrupted disc onto a
magnetic white dwarf?

\bigskip
\noindent {\bf SALVATORE ORLANDO:} Our model can be extrapolated
(with appropriate scaling) to systems in which mass accretion occurs
from a circumstellar disk, if relativistic effects can be neglected.

\end{multicols}
\end{document}